\documentclass[%
 aip,
 amsmath,amssymb,
 reprint,%
]{revtex4-1}

\usepackage{graphicx}% Include figure files
\usepackage{dcolumn}% Align table columns on decimal point
\usepackage{bm}% bold math
\usepackage[utf8]{inputenc}
\usepackage[T1]{fontenc}
\usepackage{mathptmx}
\usepackage{etoolbox}

\makeatletter
\def\@email#1#2{%
 \endgroup
 \patchcmd{\titleblock@produce}
  {\frontmatter@RRAPformat}
  {\frontmatter@RRAPformat{\produce@RRAP{*#1\href{mailto:#2}{#2}}}\frontmatter@RRAPformat}
  {}{}
}%
\makeatother
\begin{document}

\preprint{AIP/123-QED}

\title{Gadget to realize arbitrary polarization transformation on a higher order Poincar\'e sphere}
% Force line breaks with \\
\author{Sarvesh Bansal}
\email{bansalsarvesh.s@gmail.com}
\altaffiliation[Currently at ]{Dipartimento di Fisica, Università di Napoli Federico II, Complesso Universitario di Monte Sant’Angelo, Napoli 80126, Italy.}
\affiliation{Department of Physics, Indian Institute of Technology Delhi, Hauz Khas, New Delhi 110016, India}

\author{P. Senthilkumaran}
\affiliation{Optics and Photonics Centre, Indian Institute of Technology Delhi, Hauz Khas, New Delhi 110016, India}

\date{\today}% It is always \today, today,

\begin{abstract}
Designing a single element for all polarization transformations on a Poincar\'e sphere is impossible due to practical limitations and hence a combination of few standard wave-plates are used to construct a gadget. With this gadget it is possible to realize a polarization transformation between any two points on the Poincar\'e sphere.  There is no such gadget available to perform arbitrary polarization transformation on a higher order Poincar\'e sphere. We present one gadget here in which the nature of polarization transformations by its elements is a mixture of holonomic and non-holonomic, since the elements belong to two different polarization topological index spaces.
\end{abstract}

\maketitle

\section{\label{sec:level1}Introduction}
The Poincar\'e sphere serves as a pivotal tool for visualizing polarization, where each state corresponds to an unique point on the sphere. The polarization transformation by retarders is tantamount to a rotation between two arbitrary points on the Poincar\'e sphere (PS).  Considerable research has focused on the realization of a gadget that performs arbitrary polarization transformation between any two points on the Poincar\'e sphere. Such a gadget is an essential part of any polarization experiment.  This device finds applications in various domains, including quantum walks with linear optics \cite{goyal2015implementation}, measuring quantum correlations in optical systems \cite{sperling2016operational}, determining Mueller matrices for optical materials \cite{reddy2014measuring}, %Pancharatnam phase measurements \cite{chithrabhanu2016pancharatnam,loredo2009measurement}, 
and generating mixed states \cite{wei2005synthesizing,barberena2015experimental}.  In all these cases, the common objective is to achieve a unitary state transformation using the fewest possible linear optical elements. In particular, gadgets composed of multiple waveplates were initially proposed and the use of three waveplates as minimal requirement for SU(2) transformations in the PS \cite{simon1989universal,simon1990minimal} was demonstrated.  Setting aside the number of elements needed to transform one polarization state to another state, design of such gadgets is important in Polarization optics.

While the construction of PS is elegant and has proven to be extremely useful in understanding homogeneous polarization, PS cannot effectively represent inhomogeneously polarized beams with singularities. Stokes singularities, due to their spatially varying polarization distributions are depicted by simply/multiply connected regions on the PS. This limitation hinders the polarization transformations and topological analysis of such inhomogeneous polarized beams using PS. To address this limitation and depict inhomogeneous polarization, spheres homomorphic to the PS have been devised. Such constructs include the Higher Order Poincar\'e Sphere (HOPS) \cite{Holleczek:11,PhysRevLett.107.053601,Cardano:12,bansal2023stokes} and the Hybrid Order Poincar\'e Sphere (HyOPS) \cite{PhysRevA.91.023801,Arora:20}. These spheres use poles with orthogonal polarization and OAM states as eigenstates. In both HOPS and HyOPS, every other point on the sphere is given as a superposition of these chosen eigenstates. It is important to emphasize that each point on these spheres represents a beam with a spatially varying polarization distribution, unlike the PS, which represents a beam with homogeneous polarization.

In this paper, we focus on HOPS and present the design of a device which enables the conversion of a given state of polarization (SOP) at point A into an arbitrary state B on the same sphere. In this attempt, we try to accomplish the job by using the standard optical elements such as wave plates and q-plates \cite{marrucci2013q, Rubano:19}.  Such a gadget will be able to convert an initial state to another arbitrary state by tuning a few elements. In other words, starting from a given state, any point on the HOPS is reachable using this gadget.

\section{\label{sec:level2}SU(2) transformation}

The polarization transformation between spatially homogeneous SOPs is governed by the elements of the three-parameter SU(2) group, where SU stands for special unitary. These polarization transformations are represented by rotations on the PS.  The SU(2) group members are the retarders, such as quarter-wave plate (QWP) and half-wave plate (HWP).  These are homogeneous waveplates that cause the initial SOP to transform into the final SOP by rotation equal to the retardance of the waveplate on the Poincar\'e sphere.  The rotation axis lies on the equatorial plane of PS, whose orientation from the $S_1$ axis, is decided by the fast axis orientation of the retarder, where $S_i, (i=1,2,3).$ is the Stokes polarization  parameter used to construct PS.\\

Since the construction of HOPS involves phase-structured beams \cite{senthilkumaran2020phase} and the sphere represents polarization structured beams with singularities, the action of a QWP or a HWP on HOPS beams is not tantamount to rotations.  This is because these homogeneous wave plates do not satisfy the holonomy condition \cite{umar2025holonomically}, as the sphere (HOPS) and the elements (QWP/HWP) belong to two different polarization topological index spaces. Therefore, to realize arbitrary polarization transformation on a HOPS, we present here a gadget that uses elements from two different index spaces.  Each space is characterized by a sphere, a certain type of polarization elements and topologically structured optical beams.

\section{\label{sec:level3}Holonomy in polarization systems and index spaces}
With the advent of structured light and structured polarization elements, it is not possible to have a universal element that performs a unitary transformation on all polarization constructs.  There are two types of polarization transformations using retarders namely, holonomic and non-holonomic \cite{umar2025holonomically}.  In a holonomic transformation the initial, final polarization states, and the trajectory are all restricted to be on the surface of the sphere, in contrast to a non-holonomic transformation.  The condition of holonomy is honoured as long as polarization transformations are confined to a particular polarization index space. The concept of index space allows a similar mathematical formalism that is used in standard PS applicable in higher polarization topological index space optics.\\

Polarization optics that uses standard PS is termed as polarization index zero space optics.  Homogeneously polarized optical beams and elements belong to this space.  Polarization optics that uses structured beams, elements and higher order Poincar\'e sphere is termed higher topological index space optics. Therefore an element that performs SU(2) transformation in one polarization index space does not perform SU(2) transformation in another \cite{umar20252}. The HOPS meant for beams with Poincare-Hopf index $\eta$ \cite{FREUND2002251, DENNIS2002201}, and $q$-plates whose charge equal to $\eta$ are the members of same Polarization topological index space \cite{umar2025holonomically}.  The $q$-plates with any retardance angle can be fabricated and a $q$-plate with $\pi/2$ retardance is called $q$-QWP, denoted by $q^Q$ and so on.  

\begin{figure}[h!]
    \centering    \includegraphics[width=\linewidth]{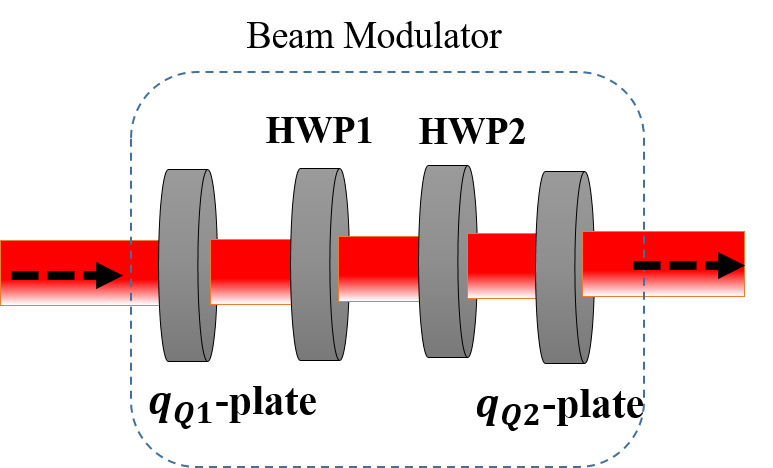}
    \caption{Proposed HOPS gadget for polarization transformation among beams represented by points on HOPS. HOPS Gadget consists of two $q^Q$-plates and two HWPs. The topological index corresponding to the two $q^Q$ plates is $q$ whereas the two HWPs belong to zero index space.}
    \label{figure1}
\end{figure}

\begin{figure*}[t!]
    \centering   
    \includegraphics[width=\linewidth]{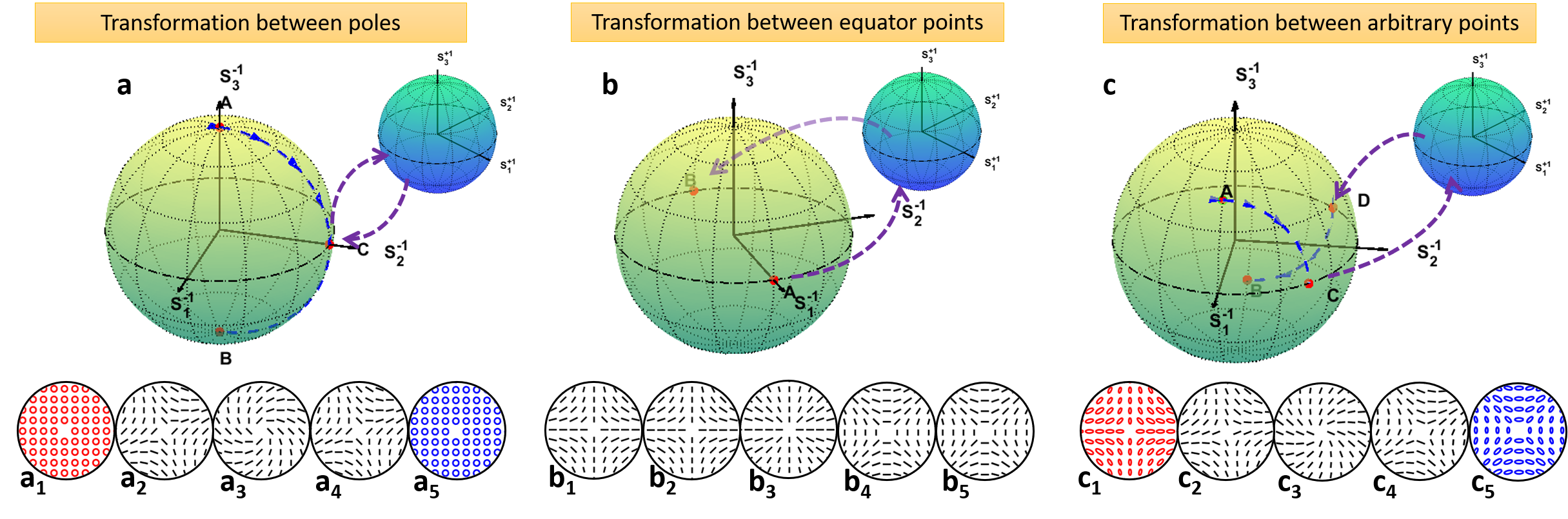}
    \caption{Polarization transformations among beams represented by points on HOPS with $\eta=-1$. (a) From the north pole $(0,\pi/2)$ to south pole $(0,-\pi/2)$, (b) among diametrically opposite point ($(0,0)$ to $(\pi/2,0)$) on the equator and (c) among any two arbitrary point A $(0,\pi/4)$ to B $(\pi,-\pi/4)$. SOP at various at various stages in our gadget is depicted below the HOPS in each case. From left to right, SOP distributions represent the incident beam ($a_1$, $b_1$ ,$c_1$), the beam after the first $q^Q$ plate ($a_2$ ,$b_2$ ,$c_2$), after first HWP ($a_3$, $b_3$, $c_3$), after the second HWP ($a_4$, $b_4$, $c_4$) and the final beam after the second $q^Q$ plate ($a_5$, $b_5$, $c_5$), respectively.  For the two $q^Q$ plates used here, $q= -1$. The blue dashed line represents the trajectory followed by the holonomic polarization transformation from $q^Q$ plate. Polarization transformation from double HWP is non-holonomic and can not be represented by trajectory on HOPS directly.  Since the HWP inverts the index of the beam, the point jumps between two spheres of $\eta=\pm1$. This is symbolically represented by inserting another (smallar) sphere of index $\eta=+1$ in these depictions.}
    \label{figure2}
\end{figure*}

\section{\label{sec:level4}Gadget made of mixed index space Optics}
Our objective is to achieve controlled polarization transformation of singular beams, which are represented by specific points on the HOPS.  In this section, we will delve into the rationale behind selecting these optical elements for effecting polarization transformations among HOPS beams.\\

The standard available optical elements such as QWP, HWP, and spatially varying wave plates are to be used to realize a gadget that works on HOPS. 
Solely relying on homogeneous waveplates proves inadequate for effecting polarization transformations on the HOPS, given that the same HOPS cannot accurately represent the transmitted beam from a homogeneous waveplate. In other words, the transformation is non-holonomic.  Likewise, an inhomogeneous waveplate falls short of providing all the necessary polarization transformations due to its fixed retardation and orientations of the local fast axis.\\

To realize polarization transformations in which both the initial and final states lie on the HOPS,  we propose a gadget composed of four elements. This gadget features a fusion of both homogeneous and inhomogeneous waveplates, specifically two HWPs and two $q^Q$-plates with matching q-values. The two HWPs belong to polarization topological index zero space, and the two q-plates belong to the same non-zero index space, which is equal to PH index of HOPS on which transformation will be performed.  The configuration entails sandwiching two HWPs between two $q^Q$-plates.  The schematic representation of this gadget is provided in Fig. \ref{figure1}. The selection and placement of these waveplates can be understood by examining the similarities between PS and HOPS.

The transformation can be understood in three parts: (a) the initial state is brought to a state represented by an equatorial point using a $q^Q$ plate by a holonomic polarization transformation, (b) a pair of HWPs transport this state to another equatorial point on the sphere by two nonholonomic transformations and (c) finally the last $q^Q$ plate transports the state holonomically to the final state.  There is index inversion \cite{pal2017polarization} every time the beam passes through a HWP.  So, the first HWP moves the state to another sphere of opposite index and the second HWP brings it back to the sphere where the beam originally belonged. 

It's crucial to highlight that the proposed HOPS gadget is effective for all HOPS except those with $\eta=+1$. This limitation arises from the radial symmetry of the $q$-plate with $q=+1$. The rotation of a $q$-plate with $q=+1$ does not alter the orientation of the fast axis, rendering it ineffective as a HOPS beam modulator. To address beam transformations on HOPS with $\eta=+1$, a slightly modified HOPS gadget can be employed. This modified setup involves inserting the HOPS gadget between a pair of HWPs.

\section{\label{sec:level5}Mathematical analysis}
\label{section7_4}
In this section, the polarization transformations are presented using Jones formalism.  Jones matrix of $q^Q$ plate is given as:
\begin{equation}
\label{eqn7_1}
    Q_q=
    \begin{bmatrix}
    \cos^2{\alpha}+i \sin^2{\alpha} & (1-i)\sin{\alpha}\cos{\alpha}\\
    (1-i)\sin{\alpha}\cos{\alpha} & i\cos^2{\alpha}+\sin^2{\alpha}
    \end{bmatrix}
\end{equation}
where $\alpha=q\theta+\delta$.  Here $\delta$ and $\theta$ is the orientation angle of $q^Q$-plate and azimuthal coordinate. Similarly, the Jones matrix for HWP oriented at angle $\beta$ is:

\begin{equation}
\label{eqn7_2}
    M_H=
    \begin{bmatrix}
    \cos{2\beta} & \sin{2\beta}\\
    \sin{2\beta} & -\cos{2\beta}
    \end{bmatrix}
\end{equation}
The combined effective waveplate for the setup is written as:
\begin{equation}
\label{eqn7_3}
   \mathbb{H}= Q_{q1}\cdot M_{H1}\cdot M_{H2}\cdot Q_{q2}
\end{equation}
 where, $Q_{q1}$, $Q_{q2}$, $M_{H1}$ and $M_{H2}$ represents the matrix for $q^Q$-plates, and HWPs oriented at angles $\delta_1$, $\delta_2$, $\beta_1$, and $\beta_2$ respectively. By solving the Eqn. \ref{eqn7_3} with the help of Eqn. \ref{eqn7_1} and Eqn. \ref{eqn7_2} the effective plate can be written as:

 \begin{equation}
\label{eqn7_4}
    \mathbb{H}=
    \begin{bmatrix}
    h_{11} & h_{12} \\ 
    h_{21} & h_{22}
    \end{bmatrix}
\end{equation}
where,
\begin{equation}
\label{eqn7_5}
    \begin{split}
        h_{11}=& \cos{(\Delta+\beta_3)}\cos{(2q\theta+\delta_3)}+i\sin{(\Delta+\beta_3)}\sin{(\Delta)}\\
        h_{12}=& \cos{(\Delta+\beta_3)}\sin{(2q\theta+\delta_3)}+i\sin{(\Delta+\beta_3)}\cos{(\Delta)}\\
        h_{21}=& \cos{(\Delta+\beta_3)}\sin{(2q\theta+\delta_3)}-i\sin{(\Delta+\beta_3)}\cos{(\Delta)}\\
        h_{22}=& -\cos{(\Delta+\beta_3)}\cos{(2q\theta+\delta_3)}+i\sin{(\Delta+\beta_3)}\sin{(\Delta)}\\ 
    \end{split}
\end{equation}
In Eqn. \ref{eqn7_5}, $\beta_3 = 2(\beta_1-\beta_2)$ signifies the relative angle between the two HWPs. Additionally, in Eqn. \ref{eqn7_5}, we have $\delta_3=\delta_2+\delta_1$ and $\Delta=\delta_2-\delta_1$. Notably, it's evident that the Jones matrix for the effective waveplate (Eqn. \ref{eqn7_4}) is independent of the absolute orientations of the HWPs, but hinges on their relative orientations to each other, as discussed earlier.

Any beam represented by points on the HOPS can be expressed as a superposition of two beams carrying equal but opposite orbital angular momentum in orthogonal spin eigenstates and written as follows:

\begin{equation}
\label{eqn7_6}
    E=A\exp(-im\theta) \Hat{e}_L+B\exp(im\theta)\Hat{e}_R
\end{equation}
where $A $ and $B$ are the complex amplitudes of superposing beams. $m$ is the magnitude of the charge of OAM carried by superposing beams. The transmitted beam $E^{'}$ from the effective plate is given as:
\begin{equation}
\label{eqn7_7}
\begin{split}
    E^{'}&=\mathbb{H}\cdot E\\
     &=A\exp(-im\theta) \mathbb{H}\cdot\, \Hat{e}_L+B\exp(im\theta)\mathbb{H}\cdot\, \Hat{e}_R
\end{split}
\end{equation}
Solving Eqn. \ref{eqn7_7} using Eqn. \ref{eqn7_5},
\begin{equation}
\label{eqn7_8}
\begin{split}
    E^{'}=&\{-A\sin(\Delta+\beta_3)\exp(-i(m\theta+\Delta))+B\cos(\Delta+\beta_3)\\
    &\exp(i(-2q\theta+m\theta-\delta_3))\}\, \Hat{e}_L + \{ A\cos(\Delta+\beta_3)\exp(i(2q\theta\\&-m\theta+\delta_3))
    + B \sin(\Delta+\beta_3)\exp(i(m\theta+\Delta))\}\, \Hat{e}_R
\end{split}
\end{equation}

It's crucial to note that the order of the $q^Q$-plate and the HOPS should align. If the $q^Q$-plate and HOPS possess different orders, the transmitted beam will assume a hybrid structure and won't be accurately represented by the same HOPS \cite{quiceno2020analysis}. Equation \ref{eqn7_8} can be further streamlined by substituting $q=m$.

 \begin{equation}
\label{eqn7_9}
\begin{split}
    E^{'}=&\{-A\sin(\Delta+\beta_3)\exp(-i\Delta)+B\cos(\Delta+\beta_3)\\
    &\exp(-i\delta_3)\}\exp(-im\theta)\, \Hat{e}_L + \{ A\cos(\Delta+\beta_3)\\
    &\exp(i\delta_3)+ B \sin(\Delta+\beta_3)\exp(i\Delta)\}\exp(im\theta)\, \Hat{e}_R
\end{split}
\end{equation}
As the transmitted beam is represented by the same HOPS then $E^{'}$ should have a similar structure to Eqn. \ref{eqn7_6} and can be written as,
\begin{equation}
\label{eqn7_10}
    E^{'}=A^{'}\exp(-im\theta) \, \Hat{e}_L+B^{'}\exp(im\theta)\, \Hat{e}_R
\end{equation}
where, $ A^{'}$ and $ B^{'}$ are the complex amplitude of the transmitted beam. By comparing the Eqn. \ref{eqn7_9} and Eqn. \ref{eqn7_10} we get,
\begin{equation}
\label{eqn7_11}
\begin{split}
    B^{'}&= A\cos(\Delta+\beta_3)\exp(i\delta_3)
    + B \sin(\Delta+\beta_3)\exp(i\Delta)\\    
    A^{'}&= -A\sin(\Delta+\beta_3)\exp(-i\Delta)+B\cos(\Delta+\beta_3)\exp(-i\delta_3)\
\end{split}
\end{equation}
\begin{equation}
\label{eqn7_12}
\begin{split}
    \chi'=\frac{\pi}{4}-\arctan{|\frac{A'}{B'}|}, \: \gamma'=\frac{1}{2}(\arg(A')-\arg(B')) 
\end{split}
\end{equation}
where, $2\chi'$ and $2\gamma'$ are latitude and longitude of the transformed beam from gadget on HOPS. Let's analyze the transformation Eqn.  with few examples. 

\subsection{Polarization transformation between poles on HOPS}

Both $q_Q$-plate are oriented horizontally ($\delta_1=\delta_2=0$). Also, HWPs are oriented parallel to each other ($\beta_1=\beta_2$). Then, $\Delta=0$, $\delta_3=0$ and $\beta_3=0$, from Eqn. \ref{eqn7_11} we get, 
\begin{equation}
\label{eqn7_13}
    A^{'}= B \quad and \quad B^{'}= A.
\end{equation} 
These results imply an interchange in the complex amplitudes of superimposed eigenstates. For example, consider illuminating the HOPS beam modulator with the specified orientation using one of the eigenstates. For instance, when illuminating with the left circular vortex beam, we have $B=0$. From Eqn. \ref{eqn7_13}, the complex amplitudes of the transmitted beams in the orthogonal eigenstates become $A^{'}=0$ and $B^{'}=A$. In simpler terms, the transmitted beam becomes an orthogonal eigenstate for this particular combination. This can be grasped by envisioning two parallel $q^Q$-plates functioning akin to a single $q^H$-plate, resulting in the interchange of complex amplitudes of eigenstates. Polarization transformation and trajectory on HOPS is depicted in Fig. \ref{figure2}(a).

\subsection{Polarization transformation between diametrically opposite points on the equator of HOPS} 

When the beam modulator is illuminated with a HOPS beam characterized by coordinates $(2\gamma,2\chi)=(0,0)$, this implies that, for the incident beam, $A=B$. If the orientation angles of the gadget are set to ($\delta_1=0,\delta_2=\pi/2$ and $\beta_3=-\pi/2$), the resulting complex amplitudes of the transmitted beam are as follows:

\begin{equation}
\label{eqn7_14}
    A^{'}= -\exp(i\pi/2) \quad and \quad B^{'}= -\exp(-i\pi/2).
\end{equation} 
From Eqn. \ref{eqn7_14}, it can be observed that the complex amplitudes of the orthogonal components in the transmitted beam possess equal magnitude and undergo a phase shift of $\pi$. Consequently, the transmitted beam is accurately represented by a point diametrically opposite on the equator of the HOPS and depicted in Fig. \ref{figure2}(b). 

\subsection{Polarization transformation between any two arbitrary points  of HOPS}

%\section{Analyzing the polarization transformations}
In the above two cases, some of the plates are positioned in such a way that the incident light is in the eigenstate of the element, so that there is no change in the state of polarization. This means that these elements are redundant in these transformations, and the presence of these elements is deliberately made ineffective to avoid removing elements of the gadget. These cases are presented to ensure that the gadget made of four elements can be used for polarization transformations between any two arbitrary points.  This means that there is no need to add or remove any element from the gadget for specific cases. Now we discuss the arbitrary transformation on HOPS. Let's take an incident beam whose SOP is represented by the coordinate ($0,\pi/4$) on HOPS with $\eta=-1$ (Fig. \ref{figure2}(c)). The final SOP which we want to achieve, is given by the coordinate ($\pi,-\pi/4$). For this two $q^Q$ plates are oriented horizontal ($\delta_1=\delta_2=0$) and the relative angle between HWPs is $\pi/8$. Then, $\Delta=0$, $\delta_3=0$ and $\beta_3=-\pi/4$, from Eqn. \ref{eqn7_11} we get, 
\begin{eqnarray}
\label{eqn7_15}
    A^{'} &=& -A \sin{(\pi/4)}+B\cos(\pi/4) \nonumber \\
    B^{'} &=& A \cos{(\pi/4)}+B\sin(\pi/4).
\end{eqnarray} 
From Eqn. \ref{eqn7_12}, we get the final coordinates as $2\chi=-\pi/4$ and $2\gamma=\pi$, same as the destination coordinates.
 
\section{Conclusion}
In recent years, extensive research has delved into the fundamental properties and applications of HOPS beams. This pursuit has led to the development of various methods for generating and detecting HOPS beams. However, up until now, no existing method has been capable of facilitating polarization transformations among HOPS beams of the same index. This article aims to bridge this gap by introducing a HOPS beam modulator, or HOPS gadget. This gadget is constructed by interleaving two homogeneous HWPs between two $q^Q$-plates. Homogeneous HWPs belong to zero-polarization index space, whereas $q^Q$ plates belong to index one space.  Through appropriate adjustments of the angles of these four components, one can effect polarization transformations between any arbitrary points on the same HOPS.  Since mixed index space elements are used, transport between arbitrary points A and B is realized through holonomic and nonholonomic polarization transformations.

\begin{acknowledgments}
Sarvesh Bansal acknowledges a research fellowship from the CSIR EMR scheme (09/086(1323)/2018-EMR-I) and P. Senthilkumaran acknowledges financial support from the Science and Engineering Research Board (SERB) India (CRG/2022/001267).
\end{acknowledgments}
\section*{Data Availability Statement}
\noindent{The data that support the findings of this study are available from the corresponding author upon reasonable request.}

%\nocite{*}
\bibliography{aipsamp}% Produces the bibliography via BibTeX.

\end{document}